# Weather Jiu-Jitsu: Climate Adaptation for the 21st Century

Qin Huang[1], Moyan Liu[1], Upmanu Lall[1,2*]

[1]School of Complex Adaptive Systems & ASU Water Institute, Arizona State University, Tempe, AZ, USA.

[2]Department of Earth and Environmental Engineering & Columbia Water Center, Columbia University, New York, NY, USA.

*Corresponding Author. Email: ulall@asu.edu

## Abstract

Extreme climate events, e.g., droughts, floods, heat waves, and freezes, are becoming more frequent and intense with severe global socio-economic impacts. Growing populations and economic activity leads to increased exposure to these events. Scaling existing physical, financial, and social infrastructure to provide resilience against these extreme events is daunting, especially as one ponders projected climate changes. Climate extremes pose a challenge even if decarbonization and geoengineering are able to regulate Earth's radiation balance.

We argue that there is an urgent need to explore a novel adaptive strategy that we call "Weather Jiu-Jitsu," which leverages the intrinsic chaotic dynamics of weather systems to subtly redirect or dissipate their destructive trajectories through precisely timed, small-energy interventions. By leveraging insights from adaptive chaos control, combined with improved observations, prediction and low-energy weather system interventions, humanity could develop a novel nature assisted global infrastructure to limit the impact of climate extremes in the 21st century.

# Main

Floods, droughts and major storms have shaped human history and the march of civilization. They continue to impact societies worldwide. In many cultures and traditional beliefs, they are attributed to divine power, signaling how gods control weather, water, and the fate of humanity[1]. Over the centuries, humans have mastered flight, tamed the atom, and arrived at the age of the Anthropocene - a human-dominated planet. Yet, control of weather and its extremes eludes us. Indeed, it is not even discussed as a priority, despite its pervasive impact on modern society. A changing climate, and its control through geoengineering and decarbonization has been part of a global debate over the last three decades, a discussion where the impact of extremes due to climate change takes center stage. These discussions consider managing the radiation balance of the planet, and do not directly address the weather and its extremes. Traditional methods to address climate extremes fail to cover catastrophic events. Losses are dominated by such events. Our perspective is that control over such weather extremes is an imperative for 21st century human society, and that it deserves to be a central focus of science and engineering.

We need to understand which weather extremes may be controllable, when, where and how. We need to develop an interdisciplinary global research agenda to control weather and avert the impact of weather extremes to complement millennia old place-based infrastructure strategies such as dams and levees, and century old strategies for heating and cooling, while addressing social and environmental concerns with both the traditional and the proposed infrastructure. Imagine harnessing the power of nature to help steer hurricanes away from land; redirect atmospheric rivers to spread their rain safely and evenly; or defuse extreme weather patterns like heat waves, deep freezes, or prolonged droughts before they ever take hold? It's a vision where we partner with Earth's own forces to create resilience, not just react to disaster. We call it Weather Jiu-Jitsu.

Climate and weather extremes pose an ever-increasing threat to human societies. Storms, floods, heat waves, droughts, tornadoes and freezes are the dominant natural hazards. Exposure to these events and their derivative events, such as fires, is growing, due in part to climate change and in part to increasing human populations and their occupancy of vulnerable areas (Supplementary Figure 1). The scale and costs of developing physical infrastructure, financial relief (e.g., insurance), and other coping programs appear prohibitive at the global scale, and traditional adaptation approaches have demonstrated the potential for adverse environmental and social outcomes. Hence, there is a need to consider a new approach to address the risk of climate extremes at a planetary scale.

A warming planet due to anthropogenic forcing of the atmosphere is the focal point of much of the climate discourse, with the intensification of the severity and duration of extremes as a major societal concern. Extreme climate events are determined largely by the dominant modes of atmospheric circulation and the associated thermodynamics and heat transport as manifest at weather time scales. The underlying equations driving these phenomena are typically nonlinear and chaotic, leading to varying and limited predictability due to sensitivity to initial conditions and perturbations. We ask whether new strategic approaches to weather modification by small perturbations could allow humanity to limit or dramatically reduce exposure to extreme climate events by nudging them and allowing the dynamics to amplify the effect. If such approaches could be developed, they would revolutionize the field of disaster management. Their technical and social implications would be far ranging, and one may require appropriate safeguards for social acceptance. This is the premise of Weather Jiu Jitsu.

Initially, we consider weather extremes, in particular those associated with mid-latitude atmospheric circulation, rather than the climatic hazards related to sea level rise or tides. This choice is motivated by the understanding of mid-latitude circulation as a nonlinear dynamical system, and that a number of climate hazards can be contemporaneously or sequentially associated with these dynamics across the globe. Given the time scales of evolution of weather anomalies, there may be opportunities for preemptive modification of their trajectories leveraging recent advances in the application of deep learning to multi-variable space-time weather prediction, via strategically timed and placed recurrent perturbations that redirect the trajectories of concern.

## Growing impacts of extreme hazards and their relation to weather dynamics

Weather and climate extremes, including droughts, floods, heatwaves, and freezes, are becoming more frequent and intense as the climate continues to warm globally[2,3]. The growing impacts are driven not only by the increasing severity of these events but also by the growing exposure of populations and assets. Global climate extremes caused an estimated $417 billion in total economic costs in 2024, of which insurance entities covered only $154 billion[4]. Hydrometeorological hazards account for 74% of losses due to natural hazards over the period from 1970 to 2019 globally[5]. The 2022 drought in Europe caused widespread socio-economic and agricultural impacts, including $26 billion in insured losses[6]. The 2021 Texas freeze in the US led to widespread power outages and economic losses estimated at $195 billion[7]. Similarly, the 2024 Mexico heatwave resulted in over 100 deaths, and the associated persistent drought caused severe water shortages and crop failures[8]. In Bangladesh, the 2022 flash flood affected over 7.2 million people, submerging vast areas and disrupting livelihoods[9].

Hurricanes, extreme rainfall, and floods pose a complex challenge given their spatio-temporal structure and impacts, that often manifest as compound extremes[10]. We note that many of such events of concern are historic, pre-dating significant climate change, and are of concern even if decarbonization efforts to arrest or reverse climate change are successful. In 1954, heavy and prolonged precipitation in central China triggered a series of flooding in the Yangtze River basin, inundating major regions of Hubei Province and resulting in over 33,000 deaths, including many from waterborne disease in the aftermath[11]. This remains one of the deadliest hydrometeorological disasters of the 20th century and predates the concern with climate change. Similarly, in the winter of 1861, California was struck by extreme winter storms, submerging large parts of the Central Valley and causing widespread destruction[12]. Looking ahead, scenario-based assessments such as the ARkStorm simulation, developed by the U.S. Geological Survey[13], highlight the escalating risk of future flood catastrophes under a warming climate. ARkStorm models the consequences of clustered Atmospheric Rivers (ARs) events that may have a recurrence interval of 250 to 1,000 years, with projections of $350 billion in physical damages and as much as $290 billion in business interruption losses in California[14]. Flood risks, both historic and emerging, are not only persistent but also intensifying as development expands, and atmospheric dynamics evolve[15,16]. Extreme weather events disrupt transportation networks, destabilize supply chains, and inflict significant economic losses across multiple sectors[17]. The social impacts of extreme weather are profound, with vulnerable populations often bearing the worst consequences[18,19]. The increasing frequency and severity of such events highlights the urgent need for adaptive strategies to mitigate risks and enhance resilience.

Persistent atmospheric blocking patterns associated with anomalous jet stream dynamics are implicated in concurrent droughts, heat waves, freezes and floods in the mid-latitudes[20,21,22,23]. Large-scale atmospheric moisture transport from the tropical oceans is often associated with catastrophic regional flooding[24], especially with recurrent tropical moisture exports associated with ARs and related phenomena[25,26,27]. ARs are among the most significant sources of climate hazards in the mid-latitudes, contributing to the increasing frequency of flooding events and severe damage, particularly along the west coasts of North America and Europe[28]. With global warming, the atmosphere's moisture-holding capacity increases which increases the potential for more frequent and intense events across diverse climate regions[29]. Large scale tropical moisture exports, including but not limited to ARs, are typically associated with fronts or eddies that subsequently interact with the mid-latitude jet stream, and are directed by the persistent high and low pressure centers and traveling waves associated with the jet stream, leading to recurrent high precipitation where they are transported to persistent low pressure centers, and drought where they are diverted away from persistent high pressure centers or atmospheric blocks (Supplementary Figure 5).

The relevance of nonlinear dynamics and chaos theory to weather dynamics was established by idealized models of such eddy-jet stream interactions developed by Saltzman[30,31], and Lorenz[32,33]. The persistent or semi-permanent features of these systems have been explored extensively since[34,35,36]. It is remarkable that so many of the climate/weather extremes of concern are associated with these types of interactions of atmospheric Rossby waves with the mid-latitude jet stream[37]. Our thesis is that Weather Jiu-Jitsu, i.e., adaptive chaos control applied to mid-latitude atmospheric dynamics may provide a 21st century approach to managing a range of extreme weather events by nudging the system trajectories. We review some of the traditional climate risk management approaches in use as a precursor to developing the argument for Weather Jiu-Jitsu.

## Traditional approaches for climate risk management

To mitigate the impacts of extreme weather events, societies have long relied on a triad of traditional risk management approaches: structural solutions (e.g., dams, levees, heating and cooling systems), financial securitization (e.g., insurance), and early warning systems. While each plays an important role, these measures are increasingly proving inadequate due to climate volatility and compounding risks. Hazard mitigation often requires synergistic responses across these 3 types of strategies. Examples of where this is effectively practiced are limited. We review each approach.

### *Physical Infrastructure*

Structural defenses including dams, levees, fire prevention systems, and cooling/heating relief centers have long been the backbone of traditional climate risk mitigation. These engineered systems aim to regulate water flow, protect populations, and preserve the continuity of critical services. In many regions, especially those with high population or asset density, these interventions have proven both effective and economically justified. Global modeling[38,39] indicates that current flood protection standards may reduce expected annual damages (EAD) in urban areas by over 90%—from potentially $1,031 billion to $94 billion annually. A growing body of vulnerability assessment research, including standardized fragility and damage curves for infrastructure such as roads, substations, and treatment plants, has also improved our understanding of how different system types respond to flooding and other hazards[40]. However, aging and inadequate dam and levee infrastructure in many countries raises concerns about the future adequacy of these systems[41,42,43].

Many dams and levees in countries like the U.S. are now over 60 years old, and for many their current condition is unknown. More than 552 U.S. dams failed due to hydrologic factors between 2000 and 2023, while experiencing rainfall with a median return period of only 5 to 10 years, and over 25,000 high-hazard dams remain at risk due to insufficient capacity to withstand rainfall events and systemic deterioration[44]. Even state-of-the-art defenses can be overwhelmed during unprecedented events that exceed their nominal design levels[45], and as conditions change, this may be more frequent. Structural defenses may also contribute to maladaptive outcomes, such as the "levee effect," where perceived safety encourages further development in flood-prone zones, thereby increasing exposure and residual risk[38,45]. Economic feasibility is another constraint: while benefit-cost ratios for large-scale infrastructure are often favorable in wealthier urban areas, they are unattractive in rural or low-income regions, where expected monetary losses are small compared to construction and maintenance costs[38,39]. As a result, protection remains uneven and reinforces existing social and spatial inequalities. Emerging alternatives such as Green and Blue-Green Infrastructure (GBGI), smart adaptive systems, and hybrid solutions face challenges of fragmented governance, limited data availability, and a lack of sustained political and financial commitment[40,46,47], may offer risk reduction only for modest extreme events.

The 2021 Texas freeze led to cascading failures of electrical and water infrastructure, with nearly $200 billion in damages, 4.5 million customers without electricity and 210 deaths, largely due to a lack of winterization of the electrical and water systems[7].  It was initially argued that the event was unprecedented. However, an analysis[48] of the climate data showed that this was the 5th such freeze in the previous 70 years. An extreme southern migration of a weather pattern that often brings Arctic air and freezing temperatures to the midwestern USA, a region

that is prepared for such events with extensive winterization (Supplementary Figure 2), was responsible. This raises the question of whether it may be more effective to invest significantly in winterization in Texas, where such events are relatively rare, or to invest in Weather Jiu-Jitsu that could defuse such an extreme event by nudging the atmospheric circulation at the right time and location?

In summary, especially for catastrophic weather extreme events, traditional physical infrastructure solutions offer only a limited solution, that is aggravated by aging infrastructure and design choices. From a layered, risk management perspective, these gaps are typically addressed by financial instruments such as insurance, catastrophe bonds, and derivatives.

### *Financial: Insurance*

Insurance mechanisms, including indemnity-based, index-based, and insurance-linked securities, are designed to pool and transfer risk across populations or financial markets. However, they can face major structural shortcomings. Indemnity-based systems are plagued by high transaction costs, delayed payouts, and low uptake, especially in flood-prone areas, due to moral hazard concerns, limited awareness, and heavy reliance on post-disaster government aid. Index-based products, while more efficient, face basis risk, wherein policyholders may not receive payouts despite actual losses, undermining trust and uptake. Insurance-linked securities like catastrophe bonds can distribute risks globally but require advanced financial literacy and strong regulation, often lacking in vulnerable regions[49,50].

Even where insurance is adopted, coverage is partial and undercapitalized. In the U.S., climate-linked insurance programs like the National Flood Insurance Program (NFIP) face chronic underfunding, with regulated premiums failing to match payout obligations. Globally, insured losses are typically a fraction of total damages, forcing governments to shoulder substantial residual costs through bailouts and ad-hoc aid, leading to unsustainable fiscal exposure. The residual risks for uninsured or underinsured populations, particularly in low-income or high-hazard zones, remain alarmingly high.

In practice, the US NFIP system is fiscally unsustainable. Despite past debt forgiveness, the NFIP remains over $20 billion in debt, with major financial stress stemming from catastrophic "hyperclustered" events, large-scale, multi-region floods driven by shared meteorological forces (e.g., hurricanes), that exceed premium recovery capacity. Just eight such events, including Katrina, Sandy, Harvey, and Ian, account for over 50% of total NFIP payouts, each causing damages greater than the annual national premium intake[51]. In parallel, recurrent, low-intensity losses, typically in high-risk areas with repeat claims, generate chronic deficits, costing the NFIP over $2 billion historically and averaging $63 million per year. These losses are not linked to rare events, but rather to frequent precipitation with return periods under 5 years[51,52]. Despite recent pricing reforms, many counties still experience expected net losses under current premiums, and uptake remains low in vulnerable communities. This results in a system where federal bailouts fill the gap, and uninsured or underinsured populations face growing residual risk, especially in low-income or coastal areas.

The NFIP could be financially viable with appropriate risk-based premiums, if the catastrophic storm/flood events that cover large areas and persist over several days were excluded. Typically, these events correspond to tropical storms, tropical storm remnants, or recurrent frontal systems associated with persistent atmospheric circulation anomalies.  In summary, physical and financial infrastructure can provide some relief from weather extremes but are overwhelmed by catastrophic, persistent and recurrent events, and addressing these risks using Weather Jiu-Jitsu would be an attractive possibility.

*Early warning systems*

Early warning systems (EWS) have advanced with the incorporation of ensemble forecasting and probabilistic models, extending lead times from 48 hours to 10 days or more. These tools enhance preparedness and can enable anticipatory action. However, their effectiveness is constrained by the varying skill and utility of forecasts. Forecasts often lack specificity and reliability, particularly for variables like regional precipitation, due to challenges in capturing ocean-atmosphere interactions and internal climate variability[53,54]. Their limited lead time restricts the window for effective early action, especially in the face of fast-evolving hazards such as floods or severe storms[55]. Additionally, low predictability for certain events like polar vortex breakdowns (Supplementary Figure 3), freezes or heat waves and drought emergence, and associated communication failures, significantly limit their efficacy. The emergence of machine learning and deep learning tools has led to improvements in the hour to week prediction time scales and is now considered as a viable alternative/augmentation to physics-based forecasts by the European Center for Medium Range Weather Forecasting (ECMWF). Utilizing these advances for Weather Jiu-Jitsu to address multi-day catastrophic weather extremes is a goal.

## Decarbonization and geoengineering

Addressing the planetary energy imbalance caused by anthropogenic greenhouse gas (GHG) emissions remains a central focus of climate policy, through strategies such as decarbonization via carbon capture, utilization, and storage (CCUS) and renewable energy deployment, and geoengineering approaches like Solar Radiation Modification (SRM), which aim to reflect sunlight and reduce global mean temperature[56]. Their ability to alter the frequency, severity, or location of extreme weather events is highly uncertain, and both spur many environmental concerns.

Decarbonization efforts have shown mixed progress. The IPCC considers full-sector mitigation necessary to meet Paris Agreement targets. While significant shifts to wind and solar are ongoing in Asia and Europe, renewable energy growth in the USA has reduced power sector emissions by only 14% since 2010[57]. On a global scale, progress remains slow. High capital costs, long infrastructure turnover cycles, and political inertia limit the speed and scale of deployment[58]. Negative emissions technologies like CCUS are viewed as critical for hard-to-abate sectors such as cement and steel, yet are under-deployed and costly, capturing only about 0.1% of current global $CO_2$ emissions[59]. Modeling suggests that, at current rates, net-zero targets may not be reached before 2070, far too late to avoid severe climate impacts[60]. Furthermore, even if full decarbonization were achieved, the legacy effects of long-lived greenhouse gases and Earth system inertia imply that extremes will persist or intensify in coming decades[56]. Studies have shown that land-atmosphere feedbacks, soil moisture dynamics, and internal climate variability play central roles in amplifying extreme events, independent of global mean temperature[61,62].

Geoengineering refers to large-scale, deliberate interventions in the Earth's climate system designed to counteract anthropogenic climate change. It is broadly categorized into Carbon Dioxide Removal (CDR) and Solar Radiation Modification (SRM). While CDR aims to reduce atmospheric $CO_2$ concentrations, SRM targets the planet's radiative balance by reflecting sunlight to cool Earth. Among SRM proposals, stratospheric aerosol injection (SAI) and marine cloud brightening (MCB) have received the most attention. Climate model simulations show that SRM could lower global mean temperatures and reduce some temperature-related extremes, including heat waves and prolonged dry spells, even under high-emissions scenarios[63,64]. However, the benefits are spatially uneven, and SRM does not address many climate-related impacts such as ocean acidification, ecological disruption, or shifts in global precipitation patterns and monsoons[60,65]. A sudden termination of SRM deployment could also cause abrupt and dangerous warming, known as "termination shock"—with potentially catastrophic social and ecological consequences[66]. Recent reviews underscore that SRM would need to be sustained for over a century even under optimistic mitigation scenarios, raising profound political and logistical challenges[65]. Additional concerns include governance and ethical issues—who controls deployment, how the risks are distributed, and whether SRM could be unilaterally implemented by powerful nations or private actors[67,68].

These emerging climate change mitigation efforts may help reverse part of the anthropogenic climate change trajectory, but their reduction of weather extremes, cascading risks, or the planetary-scale circulation dynamics driving climate extremes at specific locations of concern will likely be very limited. This underscores the need for a new paradigm—not to resist weather, but to subtly steer it: Weather Jiu-Jitsu.

## Toward Weather Jiu-Jitsu

An apt summary of the previous sections is that traditional approaches to mitigate weather and climate extremes have been successful at moderating the impact of modest extremes, keeping losses to the same percentage of GDP as the global GDP and population have grown (Supplementary Figure 4). Exposure to catastrophic extremes continues to be a factor and in the face of a changing climate motivates us to urgently explore a new 21st century infrastructure that can leverage advances in climate science, observations, artificial intelligence and related technologies to find ways to act globally to avert local and regional climate disasters. We see this as a scientific imperative that is critical irrespective of whether climate change mitigation efforts are successful or not.

To begin with we consider mid-latitude phenomenon whose physics and nonlinear dynamics are relatively well understood, and whose chaotic dynamics at the time scales of interest suggest that there is an opportunity to intervene, at least for some of the extremes that would otherwise have catastrophic impacts. The open technical questions are (a) whether appropriately timed and placed small nudges can indeed allow controls of the trajectories of weather systems of concern, (b) how these nudges would be applied in practice, and (c) how the system would be monitored and predicted to assure that the emerging trajectories are working as intended, or are updated with new nudges. These are the main elements of adaptive control.

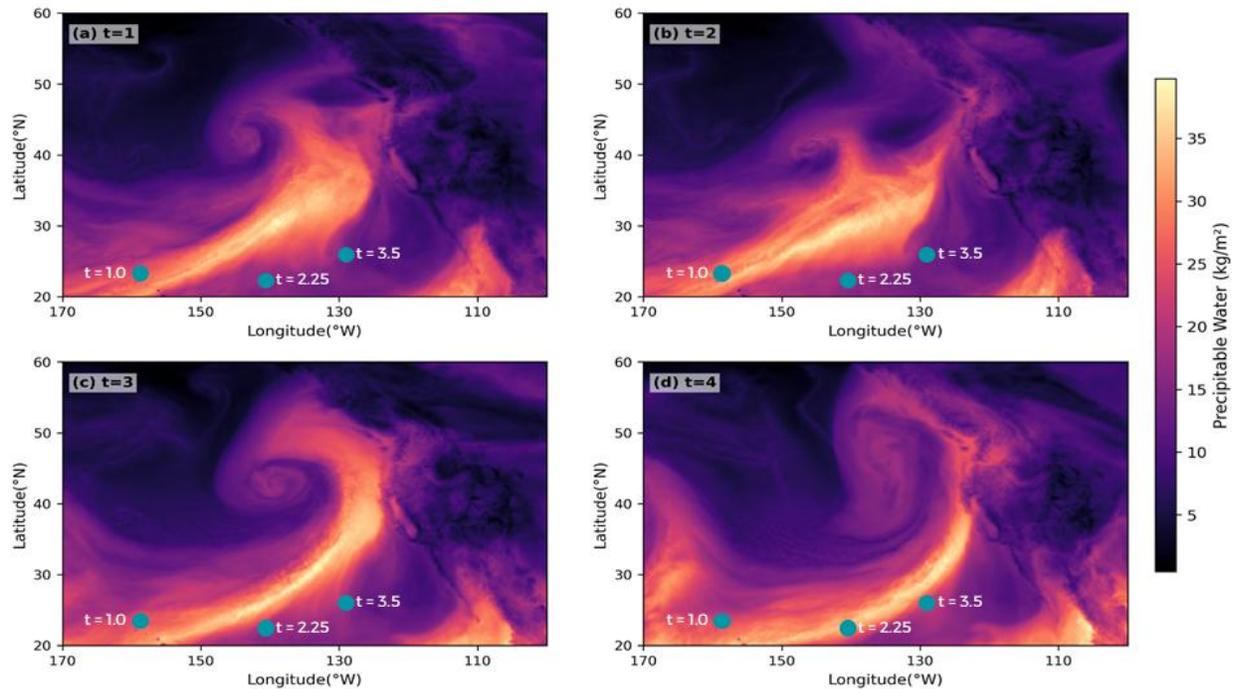

Figure 1. Conceptual illustration of an intense atmospheric river (AR) event (shown through the intensity of precipitable water (kg/m²)) approaching the western coast of North America. Panels

(a–d) correspond to snapshots on four consecutive days (t = 1 to t = 4), visualizing the evolving AR structure and trajectory. Cyan dots indicate the locations and times of small perturbations ("nudges") introduced into the system to modify the AR's path. The first dot corresponds to a nudge at t = 1.0, the second at t = 2.25, and the third at t = 3.5. This targeted sequence of low-energy interventions leads to a redistributed AR landfall precipitation footprint by t = 4, illustrating the potential to transform a high-impact landfalling precipitation that could cause a major flood into a more distributed and manageable rainfall event through adaptive chaos control. Adapted from NOAA Global Forecast System (GFS) model output[69].

The concept of *Weather Jiu-Jitsu* leverages the nonlinear and chaotic nature of the atmosphere, inspired by Edward Lorenz's seminal work on idealized systems such as the Lorenz 63 (L63) and Lorenz 84 (L84) models[32,33,70]. His work revealed two key features of atmospheric dynamics: intransitivity, or the existence of multiple coexisting flow regimes, and chaos, which refers to the extreme sensitivity of the system to initial conditions. The typical application of these ideas has been to speak of the limits of forecast predictability. However, sensitivity to initial conditions also implies that small, well-timed perturbations may be able to steer the atmosphere toward desirable regimes. This idea forms the foundation of Weather Jiu-Jitsu: a paradigm that exploits the natural instabilities and bifurcations in the climate system to gently "nudge" it away from harmful trajectories using minimal energy input.

The L63 and L84 models have been popular with dynamicists to explore attractor structure, predictability, prediction using machine learning tools, and also as means for control. These are "toy models" of convection in an asymmetrically heated rotational system (L63), and of the interaction between the atmospheric jet stream and the eddies that are coupled to it (L84). For L63, a number of chaos control strategies have been shown to be numerically and physically feasible to control regime switching and trajectory stabilization[71,72]. Recent advances have integrated deep learning to approximate system dynamics and guide interventions, while Lyapunov Exponents (LEs) which quantify sensitivity to initial conditions are used to identify where and when to apply perturbations[73,74,75]. Furthermore, researchers have successfully used Data Assimilation (DA) to estimate the system state and apply Model Predictive Control (MPC) to apply directional forcing that steers chaotic systems like L63 toward specific attractor basins[76,77,78]. Similar results are being achieved for L84, which has a more complex attractor[79], and has been explored with seasonal variations[33,80,81] in the forcings provided by the equator to pole and land-ocean temperature gradients, as well as ENSO forcings[82]. This strategy is now being explored at broader spatiotemporal scales, with efforts like Japan's Moonshot 8 project aiming to demonstrate weather control feasibility by 2050[83].

How can one transition from controlling chaos in toy mathematical models of the atmosphere to controlling atmospheric circulation in the real world with small perturbations? Which features are likely to be most or least tractable and over which space and time scales? In the mid-20th century, the USA pursued Project Stormfury[84,85,86], whose goal was to weaken hurricanes using a variety of possible weather control methods. One of the hypotheses was to modify the convective cloud properties through cloud seeding or other means, effecting a weakening of the winds and the associated rainfall. The number of actual experiments was few, and the results were deemed inconclusive as to hurricane modification, although effects of seeding were noted in at least one of the experiments. We reframe the question of hurricane modification by asking whether it could be more effective to modify the steering winds of a hurricane than to modify the power of the hurricane? A substantial literature discusses the role of the jet stream and associated wave dynamics in determining hurricane tracks, especially as the tropical cyclones enter the mid-latitudes[87,88,89,90,91]. Poor predictions of the jet stream in the vicinity of the evolving hurricane have been shown to be responsible for poor hurricane track prediction. Can the jet stream dynamics sufficiently upstream of the hurricane be adaptively perturbed in a way that favorably impacts the resulting hurricane track?

A related question is whether one could perturb the trajectories of Atmospheric Rivers, or large-scale Tropical Moisture Exports that translate into extreme regional floods? In the mid-latitudes, these features connect to the stationary and traveling waves associated with the jet stream and eddies or fronts that bring latent and sensible heat from the tropics. These are essentially the dynamics represented in the L84 model at a highly idealized level.

The atmospheric blocking patterns and associated troughs that set up persistent high- and low-pressure centers often associated with concurrent heat waves (Supplementary Video 1) / droughts or freezes / recurrent floods, are another interesting target. Recent research[92,93,94] using Lyapunov exponents as a diagnostic tool has identified that these phenomena are marked by particularly high instability or low predictability as the system transitions in or out of a block, while predictability increases as the block persists. The implication is that if a prediction system could detect an impending state transition into a blocking setting, early action on appropriate nudging could avert that situation.

To adaptively control weather extremes, we can consider two settings. The first considers conditional perturbation based on near term prediction, e.g., as a hurricane or an atmospheric river is in place, and we wish to change its trajectory with a finite set of monitored nudges. The second condition would be one where the system is regularly nudged, perhaps at sub-seasonal time scales to reduce the potential for undesirable weather regimes, e.g., the atmospheric blocking patterns that lead to persistent heat waves, freezes, droughts and floods. In this setting, one may need an identification of the latent states of the atmospheric circulation that include these regimes, and their transition probabilities to decide when to nudge and change the probabilities of regime transition. In both situations a deeper understanding of the circulation dynamics, its predictability and regime dynamics is needed. We consider low order models, like Lorenz's, to be useful for building such intuition and emerging spatio-temporal deep learning models as operationally useful for developing these ideas.

One would need to explore methods to explore the stable and unstable manifolds of the spatio-temporal dynamics to identify feasible or optimal solutions to when or where or how often to perturb and how this can be done. Given the recent advances in multiscale deep learning models of the atmosphere[95,96,97], the question is ripe for study, and we propose that an inter-disciplinary group take this on as a fundamental challenge.

The practical challenge is how to physically perturb the atmospheric circulation. This comes down to identifying feasible ways of delivering the energy of perturbation such that the flow is perturbed in the right way, i.e., the perturbation is amplified by natural dynamics. This is an open challenge. The Stormfury cloud seeding experiments demonstrated that silver iodide as well as salt provided the nucleation necessary for precipitation and hence the thermodynamic agency for changing the energetics. Laser lightning control has emerged[98] as an interesting research area, where lightning goes to a cloud level focal point created by multiple ground-based lasers, rather than to the ground, leading to a significant concentration of heating at the focal point. Other ideas as to how to accomplish perturbations with low energy inputs need to be explored and tested.

Finally, multiple algorithms for adaptive stochastic or chaos control have been applied to a number of problems[99,100], including spatio-temporal dynamics. Consequently, the literature on appropriate algorithms is accessible, and the challenge is the acquisition of and access to real time data on evolving systems and its assimilation into the state space for adaptive control. The minimum set of observables that is necessary for this exercise, and potentially available from different types of sensors, needs to be identified.

We recognize that there are numerous legal, social, and environmental challenges posed by our proposal. Indeed, on July 5, 2025, US Congresswoman Marjorie Taylor Greene said that she plans to introduce a bill to ban "weather modification" and geoengineering. However, unlike geoengineering strategies that impose abrupt or large-scale changes, Weather Jiu-Jitsu focuses on subtle, adaptive intervention, integrated with real-time forecasts and physical constraints. Direct impacts will be restricted to the event's time scale. Its potential risks may prove more manageable, and its benefits more precisely targeted than conventional geoengineering or hard infrastructure. By leveraging the atmosphere's inherent sensitivity, Weather Jiu-Jitsu seeks not to overpower nature, but to collaborate with it: nudging storm tracks, influencing jet stream meanders, and possibly mitigating compound extremes before they fully develop. We invite discussion and collaboration to develop and discuss ideas.